\documentclass[aps,preprint]{revtex4}

\usepackage{graphicx}

\usepackage{caption}
\usepackage{setspace}
\usepackage[mathscr]{euscript}
\usepackage[utf8]{inputenc}
\usepackage{amsmath}

\usepackage[utf8]{inputenc}

\usepackage{comment} 

\begin{document}

\title{Complete and Robust Magnetic Field Confinement by Superconductors in Fusion Magnets}

\author{Natanael Bort-Soldevila, Jaume Cunill-Subiranas and Alvaro Sanchez$^{*}$}

\affiliation{Departament de F\'isica, Universitat Aut\`onoma de Barcelona, 08193 Bellaterra, Barcelona, Catalonia, Spain}


\begin{abstract}

The fusion created by magnetically confined plasma is a promising clean and essentially unlimited future energy source. However, net energy generation has not been yet demonstrated in fusion experiments.
Some of the main problems hindering controlled fusion are the imperfect magnetic confinement and the associated plasma instabilities. Here, we theoretically demonstrate how to create a fully confined magnetic field with the precise three-dimensional shape required by fusion theory, using a bulk superconducting toroid with a toroidal cavity.  The combination of the properties of superconductors with the toroidal topology makes the vacuum field in the cavity volume consisting of nested flux surfaces, a condition for optimum plasma confinement. The coils creating the field, embedded in the superconducting bulk, can be chosen with very simple shapes, in contrast with the cumbersome arrangements in current experiments, and can be spared from the large magnetic forces between them. Because the field shape in the cavity is given by the boundary conditions in the superconductor surface, the system will always tend to maintain the optimum field distribution in response to instabilities or turbulence in the plasma. This field shape is preserved even when holes are drilled in the superconductor to access the plasma region from the exterior. We demonstrate by numerical simulations how a fully-confined magnetic field with the three-dimensional spatial distribution required in two of the most advanced stellarators, Large Helical Device and Wendelstein 7-X, can be exactly generated, using simple round coils as magnetic sources. We argue that state-of-the-art high-temperature superconductors already have the necessary properties to be employed to construct the bulk superconducting toroid. The present strategy can lead to optimized robust magnetic confinement and largely simplified configurations in future fusion experiments.
 
{\it $^{*}$ To whom correspondence should be addressed: alvar.sanchez@uab.cat}


\end{abstract}


\maketitle

One of the main challenges of our society is to produce clean energy in ways that do not damage the environment. Nuclear fusion of light nuclei -the energy of the Sun- is the most promising technology for a clean and safe solution for our long-term energy needs. Fusion requires the spatial confinement of plasma at high temperatures and pressures \cite{cowley,boozer,ongena}.
Tokamaks and stellarators are among the most advanced strategies to realize the fusion reaction; they are based on the confinement of thermonuclear plasma by toroidal magnetic fields \cite{ongena,helander12,xu}, because confined trajectories of magnetic fields are only possible for tori \cite{boozer}. 

To confine magnetic fields in toroidal geometries it is necessary to have topologically stable nested flux surfaces in the plasma volume \cite{boozer}. Flux surfaces are defined as those for which the magnetic field {\bf B} is parallel to them; field lines must lie on the toroidal surfaces, coming arbitrarily close to every point of that surface as the number of toroidal transversals goes to infinity \cite{boozer,helander_RPP}. The field required for plasma confinement can be realized by a rotational transform of a toroidal magnetic field so that the field has to have both toroidal and poloidal components \cite{boozer,helander_RPP}.
The toroidal field can be created, for example, by current loops in the surface of a toroid, as in tokamaks.  The required superimposed poloidal field can be 
produced by either a toroidal current induced in the plasma by an external ac field as in tokamaks, or by external coils as in stellarators \cite{stacey,helander_RPP,helander_comp}.
For the latter option, two ways of twisting the magnetic field lines exist: elongating the flux surfaces and making them rotate poloidally as one moves around the toroid or making the magnetic axis non-planar \cite{helander_comp}. Some stellarators like Large Helical Device (LHD) are based only on the first option and others, like Wendelstein 7-X (W7-X)  or TJ-II, use both \cite{helander_comp,spong,wang,weitler,gates,sengupta}.

The key to magnetic confinement of plasma, and therefore, to controlled fusion, is to generate the adequate magnetic flux surfaces in the volume containing the plasma and to maintain these magnetic surfaces in the event of plasmas instabilities or turbulences \cite{boozer,helander_RPP,ongena,hazeltine,freidberg,xanthopoulos,warmer}.
In spite of decades of intense efforts, including clever computer-helped optimizations for coil designs \cite{gates,wolf,mercier}, in current fusion experiments a magnetic configuration that is robust enough to ensure confinement in the presence of plasma instabilities has not been achieved. This is the main reason impeding the realization of the long-sought net-energy fusion reaction \cite{ongena,cowley,helander_RPP,xanthopoulos}.

In this work, we demonstrate how to perfectly confine a magnetic field in a toroidal volume, with the precise shape of the desired magnetic flux surfaces. This is achieved in a toroid made of bulk superconducting material that has a cavity carved along the toroidal direction and a set of embedded coils in the poloidal one (see Figs. 1 and 2).  
From fundamental electromagnetic theory and 3D finite-element simulations, we demonstrate that the field created by the coils is confined in the toroidal cavity and that the field distribution directly results from the shape chosen for the cavity, independently of the configuration of the coils, so that the field can be made with the exact shape of the desired flux surfaces. 
The system will tend to react against magnetic perturbations or instabilities by preserving the parallel-field boundary-condition at the superconductor surface.
The properties of superconductors and the topology make the system robustly preserve the magnetic confinement shape even when windows for accessing the plasma are incorporated into the system.

Superconductors have been already proposed to be used in fusion magnets as a set of tiles or monoliths that helped shape the magnetic field created by simple coils \cite{bromberg}. Another recent related strategy proposes using permanent magnets instead \cite{PRL_magnets}. However, in these proposals the field would be not fully confined to the desired region, and, more importantly, the field configuration is fixed, given by the spatial disposition of coils and tiles, which could not readily respond to magnetic perturbations in the plasma region. Instead, our approach considers a continuous bulk superconductor that not only fully confines the magnetic field inside the cavity volume but would react to any field modifications by inducing currents in the superconductor that will naturally preserve the boundary condition and thus the optimum magnetic shape. Our concept is reminiscent of the idea of 'magnetic molding' by a conducting shell \cite{molding}, which was never put in practice because of the impossibility of properly discretizing the shell into a set of conducting wires that is valid for all field shapes. In our proposal, the continuous superconducting surface responds to any field modification tending to restore the desired flux surface.

In general, magnetic fields emanate from the current loops and coils with a fixed spatial shape (dipolar at large distances). All current fusion experiments, and also particle-accelerator or magnetic-resonance magnets, have loops or coils as their magnetic sources, so they cannot avoid the stringent limitation of obtaining a field landscape resulting from the superposition of the field shape of the coils. Also, the fields of coils are extended all over the space, which prevents field confinement. In this work, we overcome these limitations by the combination of the properties of superconductors and the topology.

Here we regard superconductors as linear media with a magnetic permeability $\mu$ of near-zero value \cite{ZMP} so that the magnetic induction {\bf B}=0 in their interior.  This property is routinely applied to magnetic shielding; very good shielding of fields larger than 1T has been achieved by bulk high-temperature superconductors (HTS) as thin as a few millimeters \cite{roadmap,denis}. 
In general, the field of a magnet or a coil
fully enclosed inside a superconducting shell does not leak outside the volume \cite{gomory} (with the exception of some superconductor topologies such as a toroidal one, as discussed below). In contrast, from Amp\`ere's law, the field of current-carrying straight wires surrounded by a superconductor cylinder always exits the enclosure. We experimentally demonstrated in \cite{ZMP} this property for a straight wire surrounded by a HTS cylindrical tube.

\begin{figure}[t]
	\centering
	\includegraphics[width=0.9\textwidth]{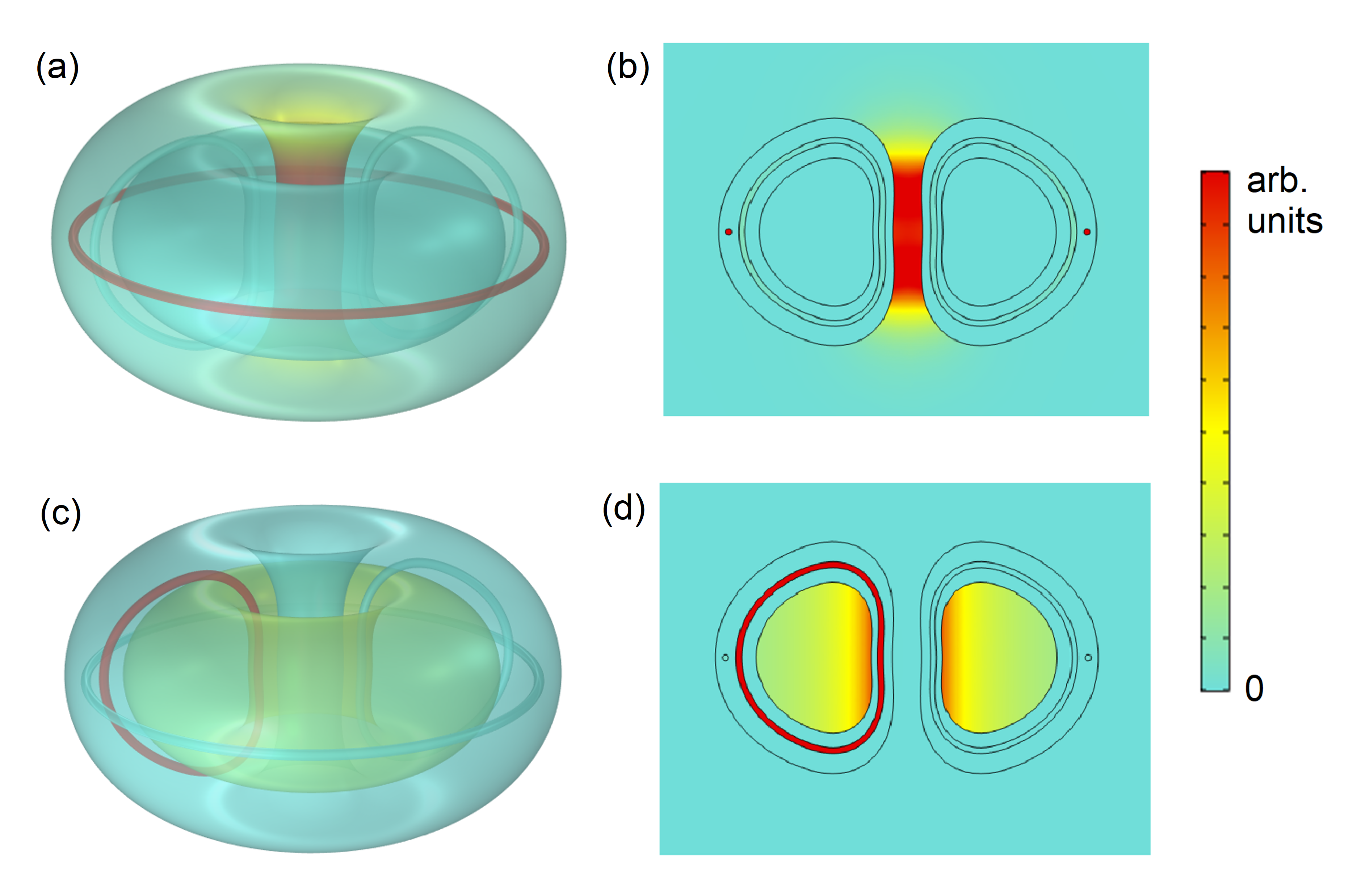}
	\caption{ 
 (a) and (b) Finite-element simulations of the 3D and 2D magnetic-field strength {\bf B} field maps, respectively, for an embedded toroidal current loop in a bulk superconductor toroid with a cavity, for which {\bf B} leaks outside the superconducting toroid. (c) and (d) 3D and 2D maps, respectively, for an embedded poloidal current loop; in this case, all the field is confined inside the toroidal cavity.}
\end{figure}

We extend these ideas to a toroidal topology. Consider a toroid made of superconducting material, containing a toroidal cavity carved with the desired shape. From magnetostatics, it can be demonstrated that the field created by a current loop in the toroidal direction exits the toroid and is mostly concentrated in the central region [see finite-element simulations in Figs. 1(a) and 1(b)]. When the loop is in the poloidal direction around the cavity, the field created is totally enclosed in the cavity, with no leak to the toroid exterior [Figs. 1(c) and 1(d);  see Supplemental Material for the demonstrations of all the properties].

In the case of poloidal currents [Figs. 1(c) and 1(d)], the magnetic boundary condition at the interface  superconductor-cavity (i. e. zero perpendicular component of {\bf B}) makes the field lines of {\bf B} exactly follow the cavity surface, which is, by definition, a flux surface \cite{boozer,helander_RPP}. Since the field is parallel to the surface on the interface superconductor-cavity, and in the absence of magnetic sources, the vacuum field in the volume in the cavity consists of nested flux surfaces. 
The field in the plasma region is totally independent of the shape of the coils generating the magnetic field, which allows the free choice of the most convenient shape for the sources (e. g. simple round coils). The requirements for optimum magnetic confinement of plasma are naturally achieved, in contrast with actual fusion experiments in which, using very complex coils arrangements \cite{helander_RPP,helander_comp}, flux surfaces are only approximately obtained and magnetic fields are not fully confined \cite{warmer,helander_comp}.

The perfect magnetic confinement is illustrated in Fig. 2. The field created by 18 current loops with the shape of the toroidal-field coils in the ITER experiment \cite{shinomura} [Fig. 2(a)] is inhomogeneous and leaks to the exterior of the coil [Figs. 2(b) and 2(c)]. If embedding these coils in a bulk superconductor [Fig. 2(d)] the field becomes fully confined in the toroid cavity, being exactly symmetric in the toroidal direction [Figs. 2(e) and 2(f)]. The total confinement and perfect toroidal symmetry is preserved even when some coils are missing and the symmetry of the sources is lost [Figs. 2(g)-(i)], demonstrating the independence of the shapes and symmetry of the sources with respect to the actual generated field. Results in Fig. 2 highlight further robustness of our superconducting strategy; flux surfaces will keep their shape, dictated only by the toroid cavity, regardless of interruptions or fluctuations in the current feeding the coils, in contrast with current fusion experiments, where changes in the coil currents can spoil the flux surface shape \cite{warmer}.

\begin{figure*}[t]
	\centering
		\includegraphics[width=0.9\textwidth]{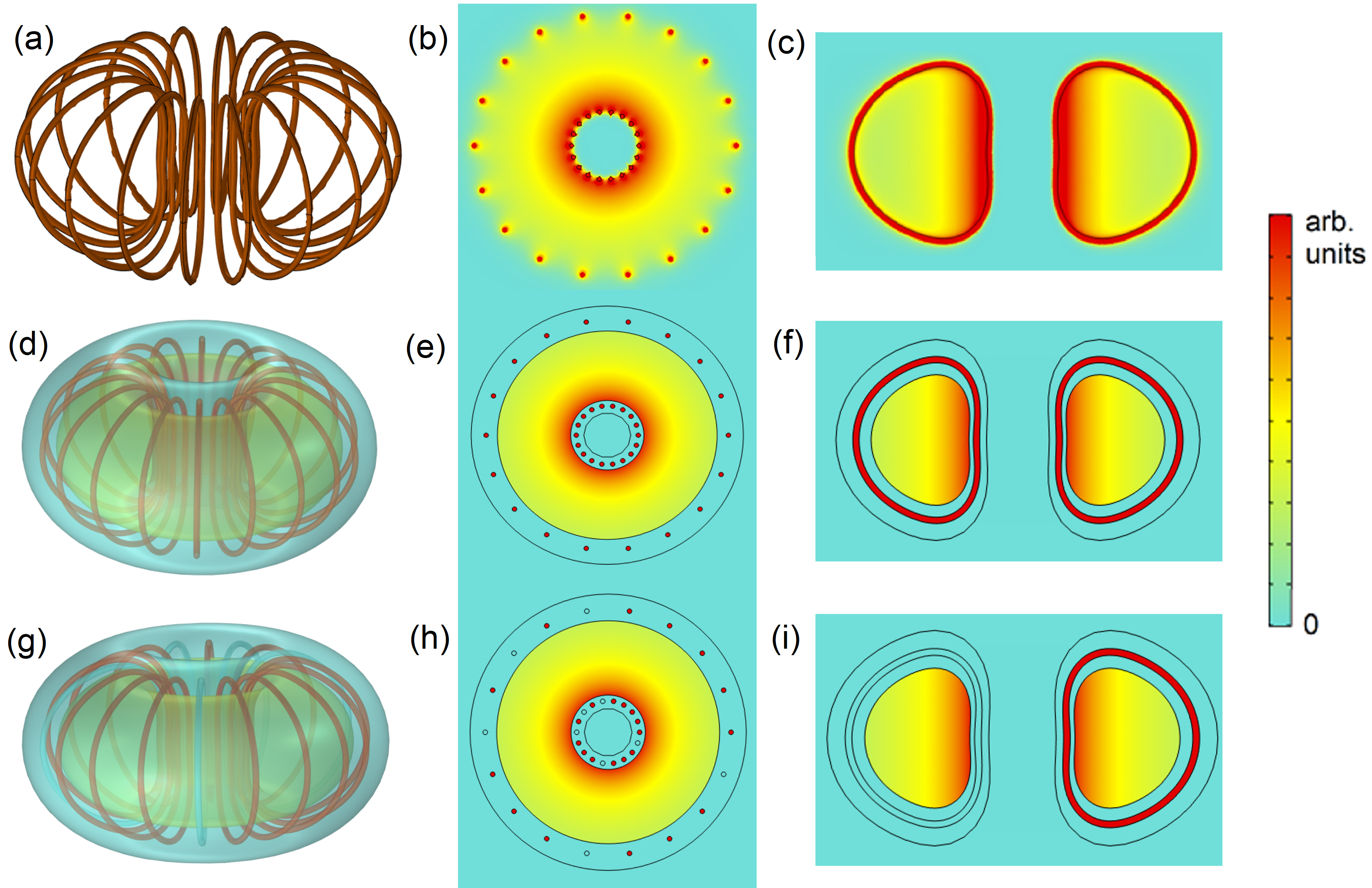}
	\caption{ 
(a), (b), and (c) Finite-element simulations of the magnetic-field strength distribution for 18 current loops similar to the ones used at ITER \cite{shinomura}. (d), (e), and (f) Field distribution of the same loops when embedded inside a bulk superconducting toroid. (g), (h), and (i) 
Field distribution when 5 out of the 18 loops are removed. The total current circulating in the previous 18 loops is now distributed in the remaining 13 loops; the obtained field is exactly the same as in (d), (e), and (f).}
\end{figure*}

Figs. 2(g)-(i) illustrate a further relevant property: the field is zero in the poloidal voids in the superconducting bulk where there are no currents. This property was experimentally demonstrated in \cite{ZMP} for parallel currents embedded in an HTS bulk cylinder with carved holes along its axis. Currents can be placed in the voids without experiencing magnetic forces from the rest of the loops; the mechanical stress is only present in the superconducting bulk, which can be made to withstand large forces, as discussed below. This may help simplify
the design of fusion magnets \cite{freidberg,shen} and can enable the use of fragile state-of-the-art HTS tapes in the coils, with a large current capacity. 

Fig. 2 shows that the toroidal magnetic field created in electromagnets like the toroidal ones in ITER [Figs. 2(a)-(c)] can be made fully confined and totally homogeneous with our approach. This will be only a partial solution for the tokamak strategy, since it is left how to generate the alternating
field that induces a current in the plasma to create the poloidal field. However, the perfect field confinement and the elimination of forces in the cables could be important advantages also to generate the toroidal fields in tokamaks. In the following we concentrate on the stellarator strategy, for which the theoretical flux surfaces will be exactly obtained from the shape of the cavity in the superconducting toroid.

\begin{figure}[t]
	\centering
		\includegraphics[width=0.75\textwidth]{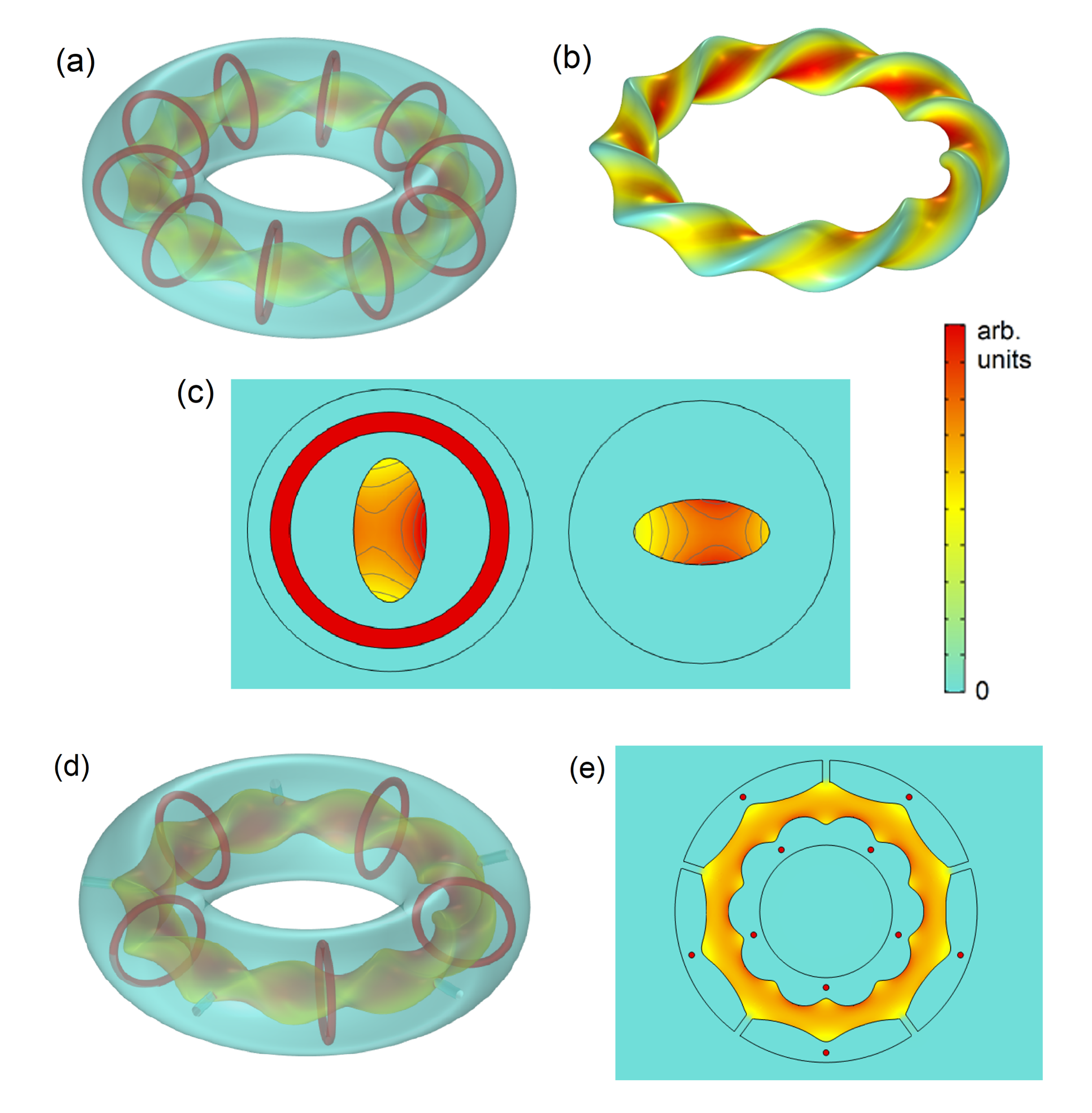}
	\caption{ 
(a) Finite-element simulations of the 3D colour map of the magnetic-field strength created by a superconducting toroid with a toroidal cavity and 10 circular loops immersed in the superconductor, corresponding to the scheme of the LHD experiment. (b) Created magnetic flux surface at the superconducting-air boundary in the cavity. (c) Field profiles at two cross-sectional cuts of the cavity at different elliptical rotations. (d) and (e) Two views of the same configuration as in (a)-(c) (with 5 currents loops instead of 10), when five symmetric holes are drilled in the superconductor, in order to have ways of access to the plasma region from the exterior.  The magnetic flux leakage to the holes is practically zero, thus preserving the flux shape in the cavity (see Supplemental Material for further data and discussion).}
\end{figure}

We numerically demonstrate how to achieve the ideal flux surfaces of two main current stellarator experiments. 
In Fig. 3(a) we show that a toroidal superconductor with a carved cavity inside with the geometry of the desired flux surface and an arbitrary number of simple round coils as magnetic sources (10 in our example), reproduces exactly the ideal flux surface in stellarators like the Large Helical Device (LHD) \cite{spong,wang}.  The magnetic field is perfectly confined in the cavity and has the theoretically required toroidal and poloidal field components embedded within the flux surface. 
The profile in the cavity cross-section corresponds to the ideal one to be obtained in LHD \cite{nakajima,saito}, with a saddle point consisting of a minimum in the direction of the minor axis and a maximum for the major axis [Fig. 3(c)] that rotates along the toroidal direction.
Interestingly, the required field profile in the cavity is maintained even when holes are drilled in the superconductor to access the plasma region from the exterior [Figs. 3(d) and (e)]. In the Supplemental Material we demonstrate that in general, thanks to the properties of superconductors, if the holes are not big compared with the cavity size and are placed with the right symmetry, the field does not leak through them except for a small exponential decay \cite{hose}, keeping in this way the shapes of the nested flux surfaces almost intact. We numerically demonstrate in the Supplemental Material [Figs. S6-S8] how the field profile in the cavity is basically unchanged even when different number of holes with different dimensions are drilled in the superconductor to access the plasma region.

\begin{figure}[t]
	\centering
		\includegraphics[width=0.45\textwidth]{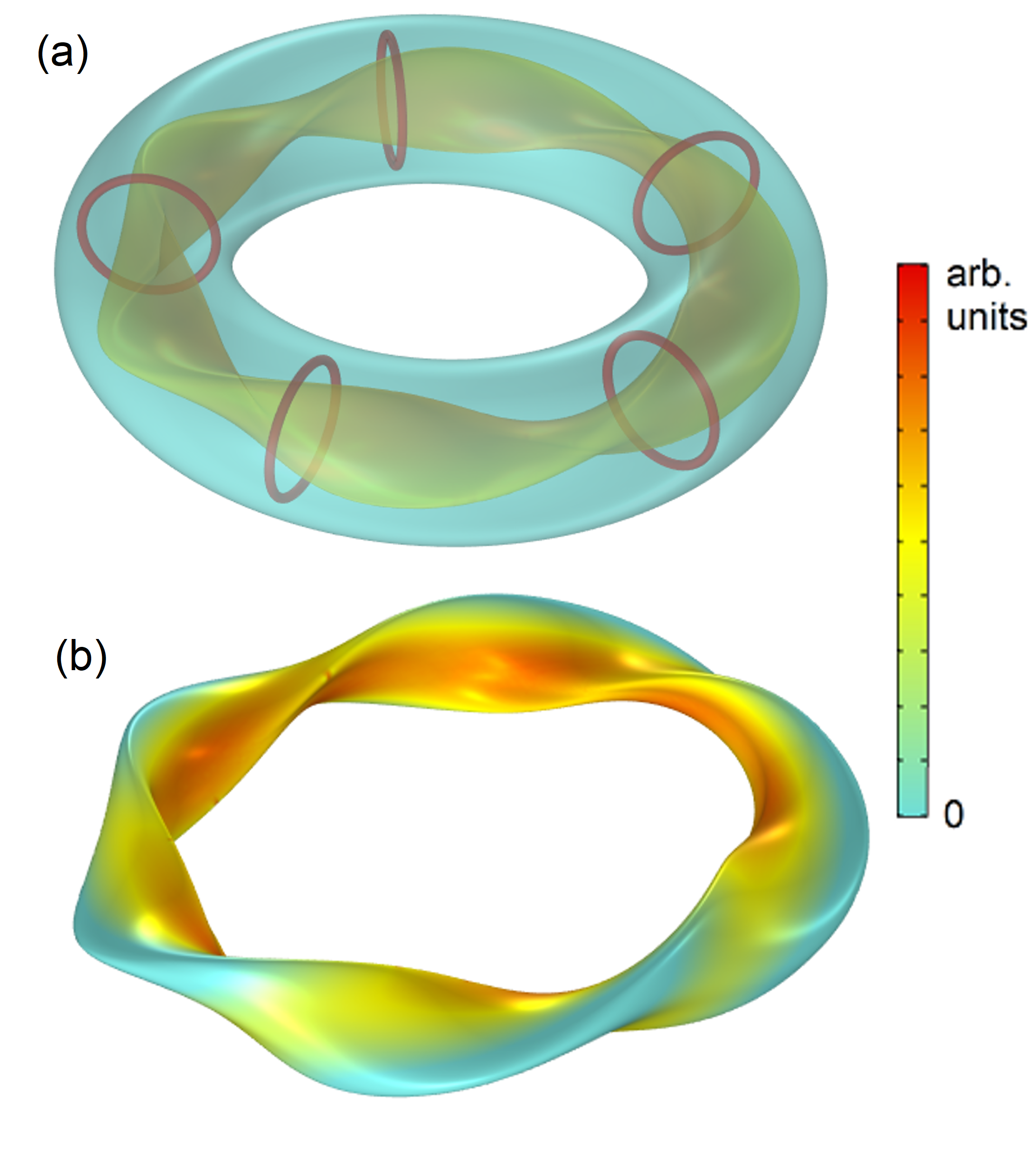}
	\caption{ 
(a) Finite-element simulations of the 3D colour map of the magnetic-field strength created by a superconducting toroid with a toroidal cavity and 5 circular loops immersed in the superconductor, corresponding to the scheme in the W7-X experiment.
(b) Created magnetic flux surface at the superconducting-air boundary in the cavity.}
\end{figure}

We further confirm our ideas in a geometry inspired by the recently constructed W7-X stellarator, which introduces an extra parameter that makes the magnetic axis non-planar \cite{helander_comp}.
Fig. 4 shows that also in this case the obtained flux surfaces correspond to the ideal surfaces theoretically designed for stellarators like W7-X \cite{weitler,gates,sengupta}.

With the results in Figs. 3 and 4, we demonstrate how a fully confined field with the 3D magnetic configuration required by the theory can be achieved with our approach, in the most advanced geometries currently developed in plasma fusion experiments and using simple round coils as sources. In contrast, the actual LHD and W7-X experiments have extremely cumbersome coil arrangements, designed after complex optimization processes \cite{gates,wolf}. They achieve only an approximation to the ideal flux surface, and the magnetic field is not confined within the flux surface but it leaks outside. Also, in our approach, the magnetic forces in the current-carrying coils are greatly reduced or even eliminated. Because the shape of the field directly results from the chosen cavity carved in the superconductor, not only the geometries corresponding to current fusion experiments, but also any other geometries suggested by theory, like a recent proposal for quasisymmetric stellarators \cite{landreman}, could be implemented following our ideas. 
These properties will make the system especially robust to magnetic fluctuations in the cavity. Magnetic instabilities will be smeared out at the outmost flux surface because any perturbation will induce superconducting currents to preserve the boundary condition of field tangential to the surface. This will act as a self-recovering mechanism. In contrast, in current experiments, magnetic disturbances interacting with the field created by fixed coils may result in runaway processes leading to loss of plasma pressure and confinement and even damage to the container walls \cite{runaway}. 

To implement our ideas in actual devices, one would need to address situations that are not considered in our simplified assumptions, such as imperfections in the flux surfaces or enabling a space to accommodate cryogenics, fuelling systems, or exhaust systems. Moreover, bulk superconducting materials should be available with the desired properties. We discuss both issues next. 

We have considered only vacuum fields for the creation of nested flux surfaces. In actual settings, transport processes, like turbulence or collisions, may lead to imperfect flux surfaces, and let the plasma escape even when there are nested surfaces. Also, fixed-boundary instabilities can occur even if the plasma boundary is fixed. The latter two issues, which are general concerns for all fusion magnets proposals, imply that having nested flux surfaces is a necessary condition for fusion, but not a sufficient one, and this has to be taken into account in any future design. Another possible practical issue is the actual location, support, and refrigeration of the superconducting bulk toroid, the crucial ingredient of our proposal. In actual fusion magnets, there is some distance between the superconducting coils and the outer plasma volume. This empty volume is needed for two main reasons: to protect the superconductors and all the external equipment from the plasma heat, which is ultimately the goal of magnetic confinement, and to accommodate divertors, tritium breeding, lithium blankets, cryogenics, etc. If all these components have no magnetic parts, they will not distort the vacuum field in the cavity. The solution we propose is to place the superconducting bulk with the embedded coils in the same volume that now occupy the complex set of coils in current stellarators. Instead of having a complex set of coils (more than 50 bizarrely twisted coils in Wendelstein 7-X) with their supports, we put in the same volume a much simpler set of round coils surrounded by the bulk superconductor with the tailored inner surface. Therefore, since it occupies the volume now taken by the superconducting coils, the cryogenics needed to cool the bulk superconductor could benefit from the same engineering solutions currently used. Also, this location ensures us that there would be basically the same distance (around 1m for Wendelstein 7-X) between the coils and the boundary of the plasma region as in current devices, with the important advantage that our strategy of using a tailored-shaped bulk superconductor makes it always tending to preserve the vacuum field shape in response to magnetic fluctuations. In contrast, these fluctuations cannot be dealt with easily with the fixed set of coils employed in the present stellarators, so they can lead to loss of plasma pressure. Since we have demonstrated that our embedded coils would experience far fewer forces than the present coils and also that some holes can be made in the bulk without affecting its properties, one can easily devise simple ways to support our coils with light structures than can pass through holes in the superconducting bulk. Finally, as discussed next, the bulk superconductor can be made in pieces so that it can be easily dissembled for maintenance tasks.

The superconductors to be used in the superconducting toroid need to operate at magnetic fields of a few Tesla, to withstand large forces, to be able to form large volumes, and to allow finely shaped cavities to be carved in them. Several families of state-of-the-art bulk HTS fulfill these requirements.
Actually, bulk superconductors are being currently used in technologies like high-performance electrical motors, superconducting bearings, flywheel energy storage, and levitation trains \cite{chapter_Ainslie}.
 Bulk superconductors of the family RE-Ba-Cu-O, where RE are rare-earth elements like yttrium or gadolinium, have shown exceptionally good superconducting properties at large fields \cite{whyte}.
Regarding mechanical properties, bulk HTS have been shown to withstand Lorentz forces arising from applied fields up to 7T \cite{Durrell 2014 SUST}. Even higher fields can be endured by reinforcement of the bulks by carbon fibre \cite{nature_murakami} or stainless steel \cite{Durrell 2014 SUST}. 
At high applied fields, some field and current will penetrate the superconductors \cite{reviewCSM}, making them depart from the ideal assumed behavior, $\mu=0$. However, the field penetration can be small for the best materials. The effective penetration depth 
at applied fields $B_{\rm a}$ is approximately $\lambda_{\rm eff} = B_{\rm a}/(\mu_0 J_{\rm C})$  \cite{reviewCSM}. 
The critical current density of RE-Ba-Cu-O HTS is larger than 10$^{8}$ A/m$^2$ at fields as large as 5T at 77K, and it increases at lower temperatures \cite{nariki_SUST}.   
Assuming  $J_{\rm C}\sim 4 \times 10^{8}$A/m$^2$, a field of 5T would correspond to a field penetration of about 1cm, which would not be a large departure from the ideal behavior in a few-meters fusion device. 
MgB$_2$ and Bi-Sr-Ca-Cu-O families also present good properties at low temperatures \cite{roadmap}.
All these superconductors are ceramics or 
metals. They can be machined with detail into desired geometries. Such finely shaped superconductors are currently in use, for example in shielding devices \cite{gozzelino} and fault-current limiters \cite{elschner,bock}. The best properties of bulk HTS have been obtained in pellets of up to tenths of centimeters; a large structure in the form of a toroid could be constructed by assembling a number of these pellets. In \cite{ZMP} we demonstrated properties analogous to those in this work using a Y-Ba-Cu-O cylindrical tube made of ten stacked disk pieces, reproducing very well the behavior theoretically predicted
for a solid piece.  Moreover, it has been demonstrated that radiation effects will not present a serious challenge for the use of high-temperature superconductors in fusion magnets \cite{weber,bartunek}.

In this work, we have introduced an approach for fusion magnets that fully confines magnetic fields in a toroidal volume carved in a bulk superconducting toroid, with a set of important properties:  (i) the magnetic field confined in the cavity has the shape of nested flux surfaces (disregarding imperfect optimizations or plasma instabilities); (ii) no field is leaked outside the cavity, which can protect the electronic equipment in the environment of the reactor; (iii) the shape of the coils energizing the magnetic fields is totally independent of the final field distributions, so that the most convenient shapes, e. g. round coils, can be employed; (iv) magnetic forces between coils can be very much reduced, allowing the use of novel fragile HTS tapes; (v) the vacuum field shape is robustly preserved even when windows are drilled to access the plasma region; (vi) the superconductor is always reacting to keep the boundary conditions at its surface, which may be helpful in the case of magnetic disturbances in the plasma.
Our approach, combining the unique properties of superconductors with topology, can be relevant in the next generation of fusion magnets to help achieve the goal of a clean and essentially unlimited source of energy.

We thank Per Helander for comments, and Gonzalo Merino, Carles Acosta, and the Port of Informacio Cientifica for help in the numerical simulations.
A. S. acknowledges funding from ICREA Academia, Generalitat de Catalunya.


\setcounter{figure}{0}
\renewcommand{\thefigure}{{S}\arabic{figure}}

\clearpage

\newpage

\vskip 2truecm

\parindent=0pt

\section*{\Large Supplemental Material: 'Complete and Robust Magnetic Field Confinement by Superconductors in Fusion Magnets'}





\vskip 4truecm

{\parindent=2truecm

{\bf 
Contents:  

\medskip







1. Methods.

2. Theoretical demonstration of the properties.

3. Drilled torus.

}

}



\newpage

\section*{{\Large\bf1. Methods}}

Numerical simulations were performed using the program COMSOL Multiphysics 5.6 with the magnetic fields (mf) physics interface, which solves the equations $\nabla \times \textbf{H}=\textbf{J}$, $\textbf{B}=\nabla\times\textbf{A}$ and $\textbf{J}=\sigma\textbf{E}$ with boundary conditions $\textbf{n}\times\textbf{A}=0$. The simulations were performed in the stationary state, using 3D spatial dimensions. The simulation mesh selected had a maximum element size of 0.08m, minimum element size of 0.002m, a maximum element growth rate of 1.3, curvature factor of 0.4 and a resolution of narrow regions of 0.7. The total calculation space was set to a 5m side length square with the relevant region of study of around 2m. For the relative permeability of the superconducting material, a value of $2\cdot10^{-7}$, in SI units, was typically used; when using other similar values no significant changes occurred in the results.

\section*{{\Large\bf2. Theoretical demonstration of the properties}}

\section*{{\large \bf Magnetostatic principles}}

In magnetostatics, the electric and magnetic fields decouple, simplifying Maxwell equations to
\begin{equation} \label{eq:maxwell1}
\nabla\times {\bf H}={\bf J}\, ,
\end{equation}
\begin{equation} \label{eq:maxwell2}
\nabla\cdot {\bf B}=0\, ,
\end{equation}
where {\bf H} is the magnetic field, {\bf B} the magnetic induction, and {\bf J} the free current density. However, in this work we refer to {\bf B} as the magnetic field, for simplicity. Due to (\ref{eq:maxwell2}), $\textbf{B}$ can always be expressed as the curl of the potential vector $\textbf{A}$,

\begin{equation}
\textbf{B}=\nabla \times \textbf{A}\, .
\end{equation}

Equation (\ref{eq:maxwell1}) can be rewritten as its integral form,
\begin{equation}
\oint_C\textbf{H}\cdot \text{d}\textbf{l} =\int_S \textbf{J} \cdot \text{d}\textbf{a}\, ,
\label{amperes law}
\end{equation}
where $\textbf{J}$ is the free current density threading a surface $S$ generated by any closed line $C$. This implies that, in the Coulomb gauge ($\nabla \cdot \textbf{A}=0$), within a spatial region containing no free currents the Laplace equation, 
\begin{equation}
\nabla^2\textbf{A}=0\, ,
\label{laplace}
\end{equation}
is fulfilled (understanding that this is valid for each component of {\bf A}).

In any interface separating two media, two boundary conditions result from (\ref{eq:maxwell1}) and (\ref{eq:maxwell2}), which are, respectively,
\begin{equation} \label{eq:contornH}
{\bf n}\times({\bf H_1}-{\bf H_2})={\bf K}\, ,
\end{equation}
\begin{equation} \label{eq:contornB}
{\bf n}\cdot({\bf B_1}-{\bf B_2})=0\, ,
\end{equation}
being {\bf K} the free surface current density and {\bf n} a unit vector perpendicular to the interface which separates regions 1 and 2. This means that the component of {\bf B} perpendicular to the surface is always continuous. 

\section*{{\large\bf Interlaced cavities}}

Consider a bulk superconducting torus with two interlaced toroidal cavities carved inside it, one empty and the second one with a current loop within it. By interlaced we understand that one threads the other, and vice versa (like the wire and the biggest toroidal cavity in Fig. 1(c,d)). In the following, we demonstrate that the field in the empty cavity has the shape of a set of nested magnetic flux surfaces.

The magnetic field at the surface of the inner cavity will be always parallel to that surface, since (\ref{eq:contornB}) leads to
\begin{equation}
\textbf{n}\cdot \textbf{B}_{\bf SC}-\textbf{n}\cdot \textbf{B}_{\bf cavity}=0\,\,\, \Longrightarrow\,\,\, 	\textbf{n}\cdot \textbf{B}_{\bf cavity}=0\, ,
\label{specific_bounadry}
\end{equation}
because $\textbf{B}_{\bf SC}=0$. 
This forces the magnetic field to be contained in a flux surface with the carved shape. This particular flux surface is called the boundary flux surface. Taking into account that there are no magnetic poles or currents inside the cavity, Laplace's equation (\ref{laplace}) is fulfilled, so, fixed the boundary conditions, a unique solution exists. Therefore, there will be magnetic flux surfaces remaining one inside the other surrounding what is called the magnetic axis. In fusion, these nested flux surfaces are needed to perfectly confine the particles inside the reactor and, neglecting the field of the plasma currents, they can be obtained exactly by using our approach.

\begin{figure}[ht]
\centering
\includegraphics[scale=0.4]{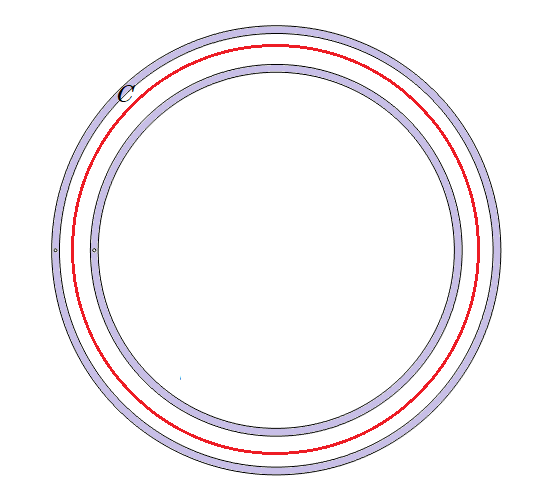}
\caption{Horizontal cross-section of a torus with a toroidal cavity and an interlaced embedded perpendicular current wire (left of the torus). A red curve $C$ is drawn centered at the center of the torus to perform Ampère's calculation.}
\label{amper}
\end{figure}

To demonstrate that the magnetic field $\textbf{B}=\mu\textbf{H}$ inside the cavity is not zero,  Ampère's law (\ref{amperes law}) is used in a particular example of a circular toroidal cavity, with one loop of current $I$ embedded in the SC. A cut in the horizontal plane of this hollow torus is plotted in Fig. \ref{amper} where a red curve $C$ is traced inside the cavity. This curve can be drawn as a circle of radius $\rho$ centered at the center of the torus. It is always parallel to both sides of the walls of the cavity, so $\textbf{B}$ will always be tangential to the curve $C$. The strength of $\textbf{B}$ should be constant all along the line, due to the symmetry of the problem. Then,

\begin{equation}
\oint_C\textbf{B}\cdot \text{d}\textbf{l}=|\textbf{B}|\oint_C \text{d}l=|\textbf{B}|2\pi \rho =\mu \int_S \textbf{J}\cdot \text{d}\textbf{a}\, .
\end{equation}
Since the surface $S$ created by curve $C$ has a total net current of $I$ threading it, the magnetic field inside the cavity is not zero and has a value 

\begin{equation}
|\textbf{B}|=\frac{\mu I}{2\pi \rho}\, .
\end{equation}

For any other cavity shape and any number of current loops embedded in the SC, the magnetic flux will not be zero, as long as the total net current $I$ of all the loops has a non-zero value.

\section*{{\large\bf Non-interlaced cavities}}

On the other hand, if the cavity and the embedded current wire are not interlaced (for example, like the current loop and the biggest toroidal cavity of Fig. 1(a,b)), $\textbf{B}$ will not permeate into the cavity (${\bf B}_{\bf cavity}=0$).  
Now we demonstrate that the field within the cavity is zero.

From the mathematical property

\begin{equation}
\oint_C\textbf{B} \cdot \text{d}\textbf{l}=0\, 
, \, \forall C\subset \mathscr{R}\,\,\, \Longleftrightarrow \,\,\, \textbf{B}_{\mathscr{R}}=0 \, ,
\end{equation}
where $\mathscr{R}$ is the entire closed spatial region in which $\textbf{B}$ could be non-zero (i.e. $ \mu \neq 0$, the cavity in our case), since there are no free currents threading a surface $S$ generated by any possible closed line $C$ inside the cavity, the magnetic field $\textbf{B}$ will be always zero at every point of the cavity.

This zero-value of the magnetic field leads to an absence of forces between current cables placed at different non-interlaced cavities, as discussed in the main manuscript.

\section*{{\large\bf Around the torus hole}}

\begin{figure}[ht]
\centering
\includegraphics[scale=0.495]{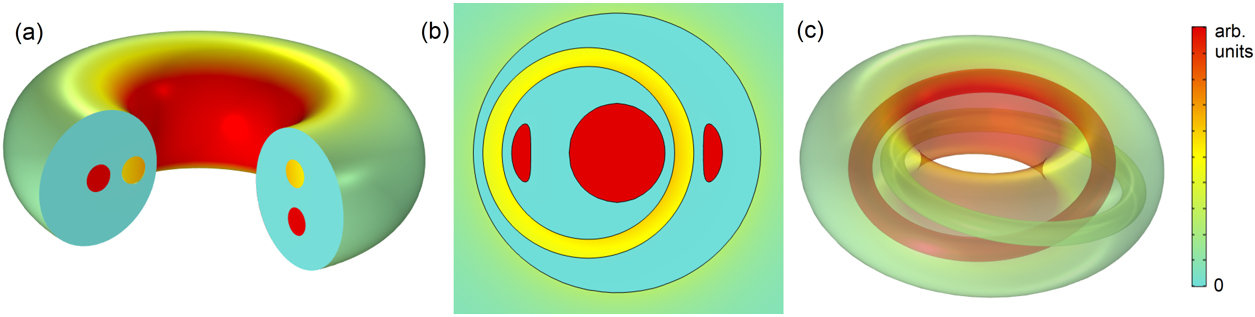}
\caption{(a), (b), and (c), Finite-element calculations of the magnetic field strength $|\textbf{B}|$ for a superconducting torus with two interlaced cavities, one with a current wire and another without it, both around the torus hole. The magnetic field penetrates the empty cavity and outside the SC torus.}
\label{intertwindescav}
\end{figure}

Another interesting property takes place when interlacing the two cavities (one containing the current wire and the other empty) around the torus hole of the SC, as shown in Fig. \ref{intertwindescav}. The magnetic field of the current wire can permeate into the other cavity and outside the SC torus. In this case, the magnetic flux also permeates outside the SC because the current wire is embedded around the SC torus hole.

This particular phenomenon allows us to modify the magnetic field inside the cavity and outside the torus simultaneously. Applied to toroidal or other kind of topologies, it can lead to novel ways to shape the magnetic field and new applications.

\section*{{\Large\bf3. Drilled torus}}

In actual fusion magnets, some holes would be needed to access the inner cavity of the torus. These holes could be drilled into the cavity by carving a tunnel connecting the outside of the SC torus and the inner cavity. Depending on where they are drilled, different outcomes of the magnetic field of the system can occur.

\subsection*{\large \bf  Field leakage}

\begin{figure}[ht]
\centering
\includegraphics[scale=0.48]{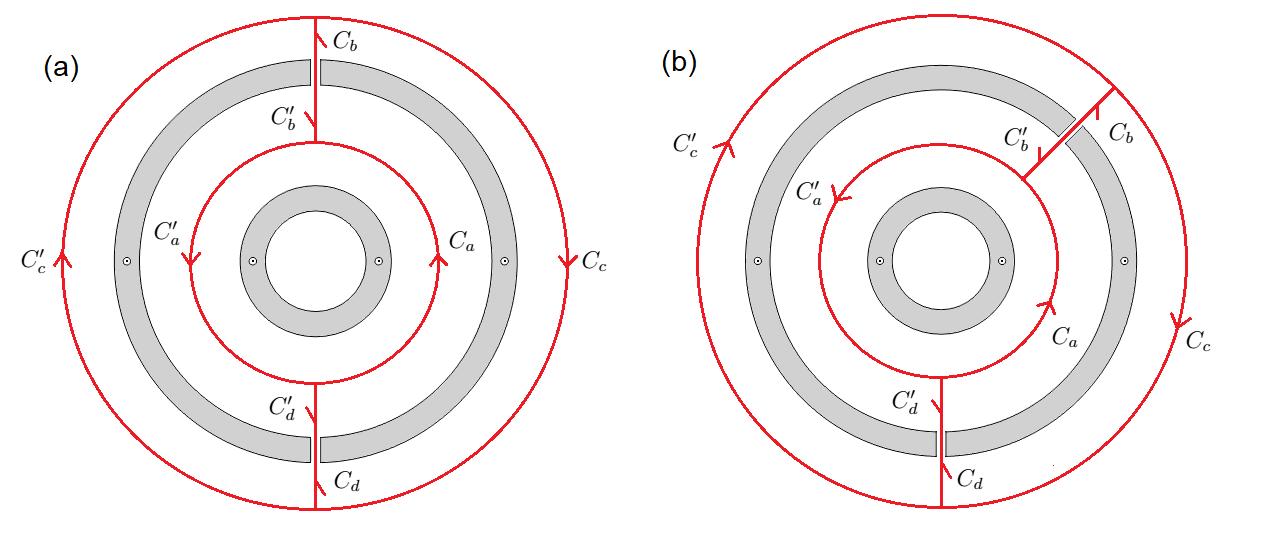}
\caption{(a), Horizontal cross-section of a symmetrical drilled torus. Two red closed curves $C$ and $C'$ are drawn to perform Ampère's demonstration. (b), Horizontal cross-section of a clearly asymmetrical drilled torus. The red curves $C$ and $C'$ are drawn too.}
\label{fig:amper_holes}
\end{figure}

Here we demonstrate the conditions needed for no magnetic-field leakage through the holes. If the system of coils and holes is as the one in Fig. \ref{fig:amper_holes}(a), where there is symmetry between coils, holes, and intensities around the coils, the magnetic field along the drilled holes will decay exponentially on first order, and the magnetic field inside the cavity will maintain the axisymmetric rotational symmetry. This axisymmetry is slightly locally broken at nearby distances to the hole, but without perturbing the global magnetic field. In such conditions, there is no major breaking of the magnetic confinement.

To demonstrate this effect we consider a simplified system, shown in Fig. \ref{fig:amper_holes}(a), with two coils (with the same current) and two holes placed at symmetric positions. In this symmetric case, the magnetic field inside the cavity would be axisymmetric, since any invariant transformation of the system has the same solution for the magnetic landscape. For example, a 180º rotation around the central torus axis remains the system unchanged and thus the same solution must result. The hole dimensions are also an important factor for the uniformity of the field in the cavity. The magnetic field decays across the hole reminding us of the field decay due to a dipole in a superconductor hole (see reference [31] of the main manuscript), where the magnetic field decays exponentially at large distances. To have an exponential decay of the magnetic field across the hole in our system, it must have a longitude-diameter quotient as large as possible. Another factor determining how local the field is distorted around the hole is the quotient between the cavity diameter and the hole diameter. The bigger this quotient is, the more local this effect will be and the less distorted the magnetic field lines will be near the hole.

If this rotational symmetry is broken, for example, by moving one hole like in Fig. \ref{fig:amper_holes}(b), the magnetic field inside the cavity will lose its axisymmetry, and the field will leak outside the cavity. This phenomenon can be understood using Ampère's law \eqref{amperes law}. Considering the two closed curves
\begin{equation*}
\begin{split}
C=C_\text{a}+C_\text{b}+C_\text{c}+C_\text{d}\; , \\
C'=C'_\text{a}+C'_\text{b}+C'_\text{c}+C'_\text{d}\; ,
\end{split}
\end{equation*}
drawn in Fig. \ref{fig:amper_holes}, we define the line integrals for the magnetic field along these lines as
\begin{equation}
\begin{split}
\mathscr{I}=\int_{C_\text{a}}\textbf{B}_\text{a} \text{d} \textbf{l}+\int_{C_\text{b}}\textbf{B}_\text{b} \text{d} \textbf{l}+\int_{C_\text{c}}\textbf{B}_\text{c} \text{d} \textbf{l}+\int_{C_\text{d}}\textbf{B}_\text{d} \text{d} \textbf{l}=\mathscr{I}_\text{a}+\mathscr{I}_\text{b}+\mathscr{I}_\text{c}+\mathscr{I}_\text{d}\; , \\
\mathscr{I}'=\int_{C'_\text{a}}\textbf{B}'_\text{a} \text{d} \textbf{l}+\int_{C'_\text{b}}\textbf{B}'_\text{b} \text{d} \textbf{l}+\int_{C'_\text{c}}\textbf{B}'_\text{c} \text{d} \textbf{l}+\int_{C'_\text{d}}\textbf{B}'_\text{d} \text{d} \textbf{l}=\mathscr{I}'_\text{a}+\mathscr{I}'_\text{b}+\mathscr{I}'_\text{c}+\mathscr{I}'_\text{d}\; .
\end{split}
\end{equation}
Therefore, by using Ampère's law \eqref{amperes law} and considering both coils with the same intensity $I$ we can obtain this equation system
\begin{equation}
\begin{cases}
\mathscr{I}_\text{a}+\mathscr{I}_\text{b}+\mathscr{I}_\text{c}+\mathscr{I}_\text{d}=\mu I \\
\mathscr{I}'_\text{a}+\mathscr{I}'_\text{b}+\mathscr{I}'_\text{c}+\mathscr{I}'_\text{d}=\mu I \\
\mathscr{I}_\text{b}+\mathscr{I}'_\text{b}=0 \\
\mathscr{I}_\text{d}+\mathscr{I}'_\text{d}=0 \\
\mathscr{I}_\text{a}+\mathscr{I}'_\text{a}=2\mu I \\
\mathscr{I}_\text{c}+\mathscr{I}'_\text{c}=0 
\end{cases}
\Longrightarrow \;
\begin{cases}
\mathscr{I}_\text{a}=\mu I-\mathscr{I}_\text{b}-\mathscr{I}_\text{c}-\mathscr{I}_\text{d}=\mu I-K \\
\mathscr{I}'_\text{a}=\mu I+\mathscr{I}_\text{b}+\mathscr{I}_\text{c}+\mathscr{I}_\text{d}=\mu I+K \\
\mathscr{I}'_\text{b}=-\mathscr{I}_\text{b} \\
\mathscr{I}'_\text{c}=-\mathscr{I}_\text{c} \\
\mathscr{I}'_\text{d}=-\mathscr{I}_\text{d}
\end{cases}
\; ,
\end{equation}
where $K\equiv\mathscr{I}_\text{b}+\mathscr{I}_\text{c}+\mathscr{I}_\text{d}$. If we consider $K=0 \Rightarrow \mathscr{I}_\text{a}=\mathscr{I}'_\text{a}$, we note that the field cannot leak outside the torus because $K=\mathscr{I}_\text{b}+\mathscr{I}_\text{c}+\mathscr{I}_\text{d}=0\Rightarrow \mathscr{I}_\text{b}=\mathscr{I}_\text{c}=\mathscr{I}_\text{d}=0$, since the field must go always in the same direction along these curves (the integrals have the same sign), hence $\textbf{B}_\text{b}=\textbf{B}_\text{c}=\textbf{B}_\text{d}=0$ for our system ($\nabla\cdot\textbf{B}=0$). Therefore, the field along $C_\text{a}$ and $C'_\text{a}$ could be equal if both lines have the same length (case shown in Fig. \ref{fig:amper_holes}(a), previously demonstrated). If they do not have the same length, the magnetic fields $\textbf{B}_\text{a}$ and $\textbf{B}'_\text{a}$ should be different but this cannot happen since the field does not leak the cavity. With this we demonstrated that having asymmetric holes (Fig. \ref{fig:amper_holes}(b)) implies $K\neq0$, since initial premise is breached, so field lines will leak the cavity by one hole and reenter by the other resulting in different values for $\textbf{B}_\text{a}$ and $\textbf{B}'_\text{a}$.

\begin{figure}[ht]
\centering
\includegraphics[scale=0.55]{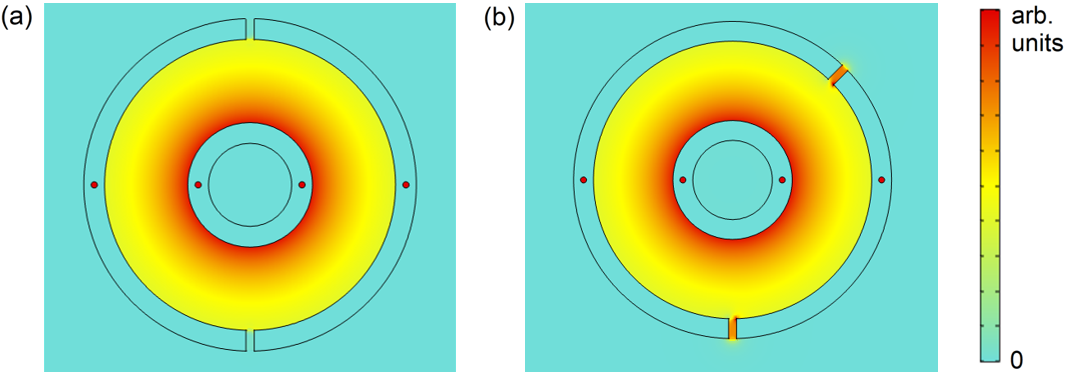}
\caption{Magnetic field strength output. (a), Symmetrical-drilled torus shown in Fig. \ref{fig:amper_holes}(a). (b), Asymmetrical-drilled torus shown in Fig. \ref{fig:amper_holes}(b). It is important to note that in (a) there is no field leakage while in (b) it leaks through the holes.}
\label{fig:amper_holes_results}
\end{figure}

From these results, we can generalize the conditions for which the magnetic field does not leak out of the cavity, and consequently, this field does not lose its axisymmetry (as long as the holes are thin and long enough, as discussed above). In general, this happens for a system in which any Ampère's curves analogous to the ones in Fig. \ref{fig:amper_holes} have the same $\frac{I}{\Delta \theta}$ ratio (being $I$ the intensity current threading the Ampère's curve and $\Delta \theta$ the toroidal angle defined by the sector limited by the two holes through which this curve passes). It is interesting to note that this solution, constant $\frac{I}{\Delta \theta}$, is independent of the number and position of the coils.

This leads to different possibilities in which the field does not leak, such as placing several holes in a transverse plane (any vertical plane defining one cross-sectional cut at a specific toroidal rotation, $\theta$), among others.

In practice, the robustness of the preservation of the field in the cavity is maintained even though these conditions are not achieved exactly. For the ideal cases in which the field does not exit, small departures from these conditions of symmetry (e. g. small variations in the positions of the holes, fluctuations in the currents of the coils) do not cause a destabilization of the system, since to obtain clearly unstable situations a great rupture of the symmetry conditions would be necessary. In fact, very asymmetric systems such as the one in Fig. \ref{fig:amper_holes_results}(b) only cause variations upto 1.3\% in the $|\textbf{B}_\text{a}|$ and $|\textbf{B}'_\text{a}|$ values (where the cavity-hole diameters quotient is 10 and the longitude-diameter quotient of the holes is 2.5). Moreover, even in very asymmetric cases, field losses to the outside and the modification of the cavity field can be compensated by properly correcting and adjusting the currents in the different coils of the system.

\subsection*{\large \bf  Implementation in fusion magnets}

\begin{figure}[ht]
\centering
\includegraphics[scale=0.143]{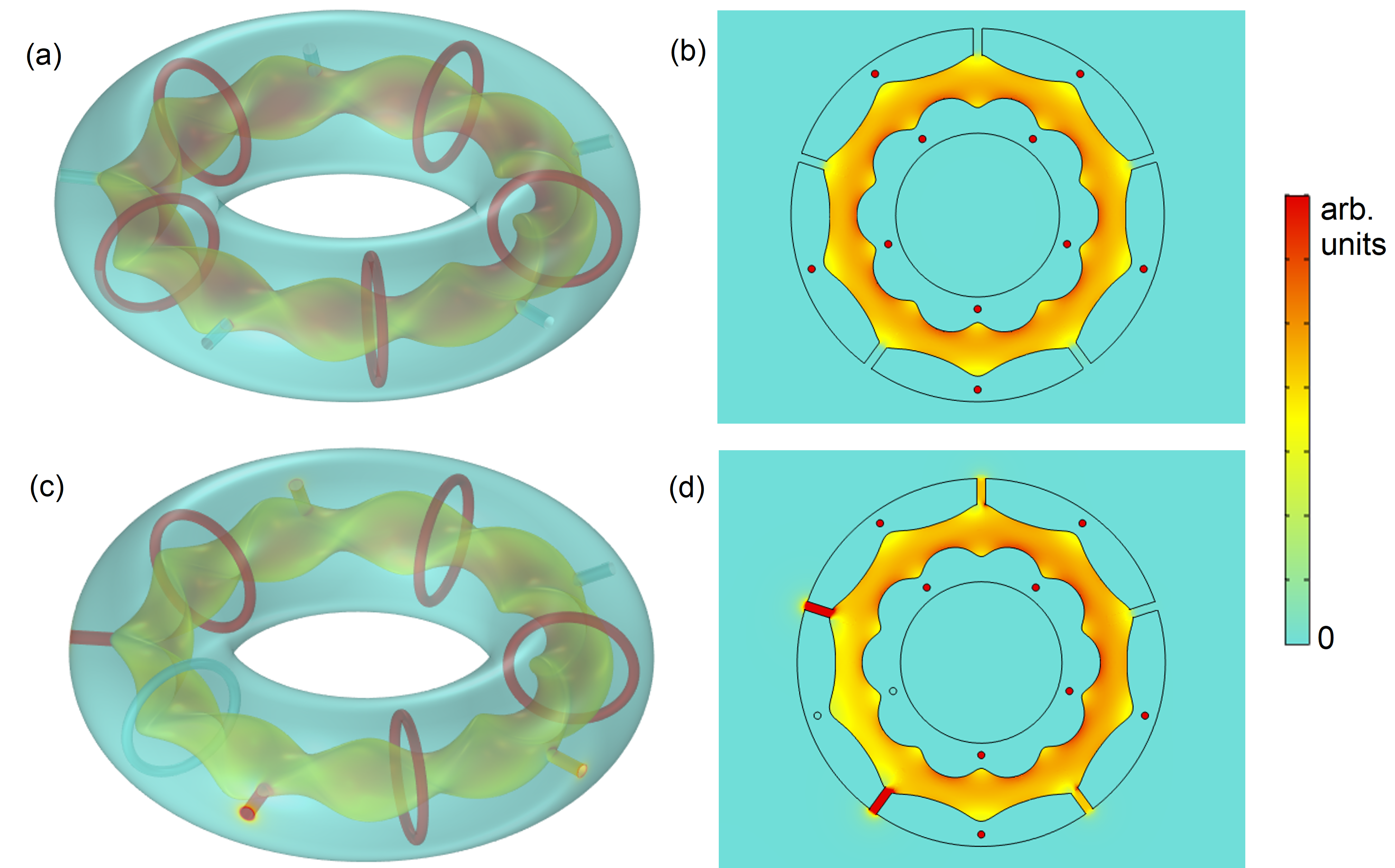}
\caption{Finite-element simulations of the 3D color map of the magnetic-field strength created by a superconducting toroid with a toroidal cavity, in which five holes to access the cavity from the exterior have been drilled. The configuration corresponds to the LHD experiment, as in Fig. 3. It can be observed that when holes are properly distributed as in (a), the magnetic flux leakage to the holes is practically zero, thus preserving the flux shape in the cavity. When the holes are not symmetric, as in (c), some flux is lost and the field in the cavity is modified. (b) and (d) are cross-sectional views of (a) and (c), respectively.}
\label{fig:LHD_forats}
\end{figure}

As an example of the application of our ideas to fusion magnets, we consider the LHD configuration, where the inner toroidal cavity undergoes a rotational transform and the field has both toroidal and poloidal components. The previously described properties are kept if the above symmetry condition is maintained, as shown in Fig. \ref{fig:LHD_forats}. This would be true also for any regular cavity shape.

\begin{figure}[ht]
\centering
\includegraphics[scale=0.25]{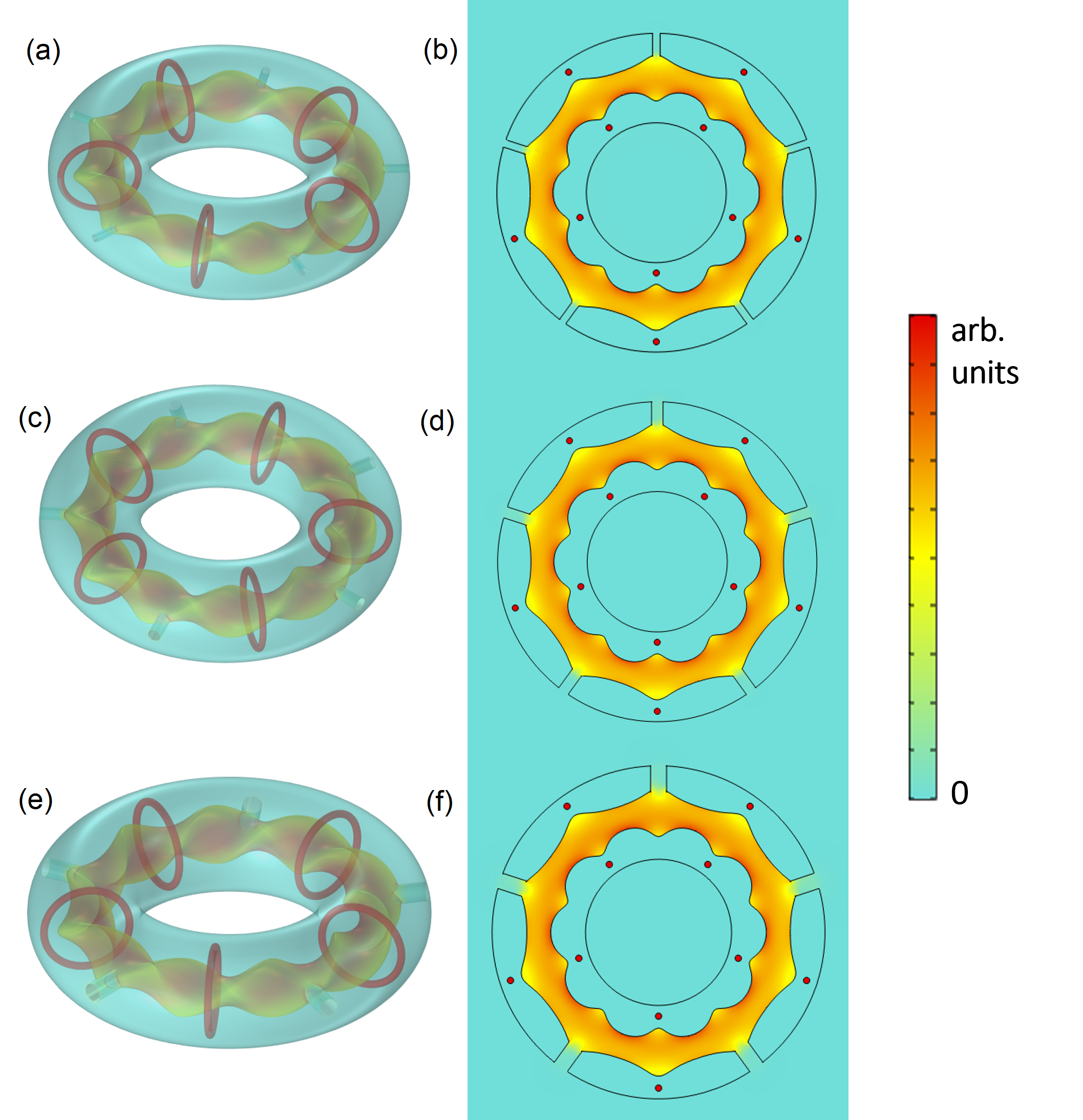}
\caption{Finite-element simulations of the 3D color map of the magnetic-field strength created by a superconducting toroid with a toroidal cavity, in which five holes had been drilled to access the cavity. Different hole radius have been drilled into (a), (c) and (e) with radii 0.05m, 0.075m and 0.1m respectively. It is important to remark the elliptical cavity has a semi-minor axis length of 0.15m and a semi-major axis length of 0.3m. (b), (d) and (f) are horizontal cut planes of (a), (c) and (e) respectively.  It can be observed that \textbf{B} decays very quickly along the hole and that local perturbations in the hole entrance are small.}
\label{fig:LHD_hole_size}
\end{figure}

When taking into account the hole size it can be seen from Fig. \ref{fig:LHD_hole_size} that the hole radius affects just locally to the magnetic field. Even for holes of the order of the cavity size, Fig. \ref{fig:LHD_hole_size} (e) and (f), perturbations remain local and without affecting the cavity field globally.

\begin{figure}[ht]
\centering
\includegraphics[scale=0.15]{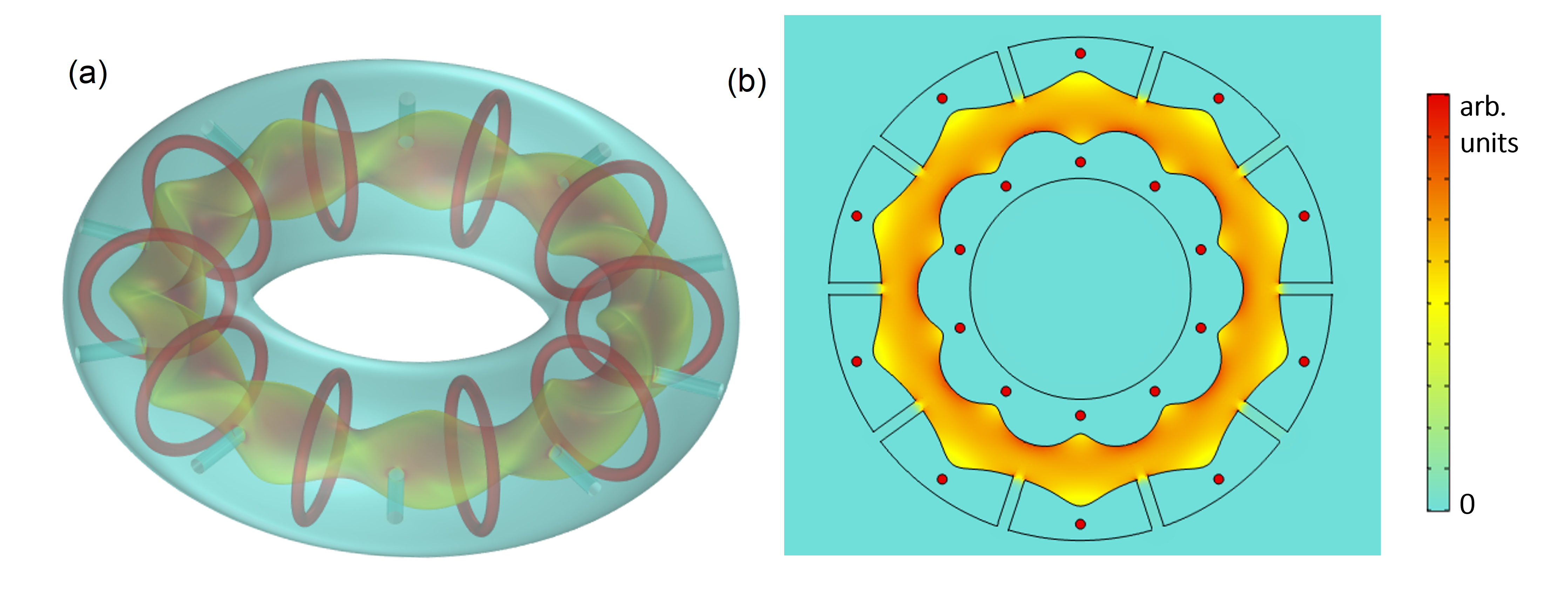}
\caption{Finite-element simulations of the 3D color map of the magnetic-field strength created by a superconducting toroid with a toroidal cavity, in which ten holes had been drilled to access the cavity. The number of current wires around the cavity is also ten. It can be observed that in the same way as for five holes and five currents no \textbf{B} threads outside the cavity. (b) is a horizontal cut plane of (a).}
\label{fig:LHD_number}
\end{figure}

The number of holes can be limited by maintaining constant the previously-mentioned quotient, $\frac{I}{\Delta\theta}$. Therefore, one possibility of increasing the number of holes is increasing the number of coils. In Fig. \ref{fig:LHD_number} we show an example, by doubling the number of coils and holes. Despite doubling the number of holes we see that the cavity field is maintained and there is no field leakage.

\begin{figure}[ht]
\centering
\includegraphics[scale=0.15]{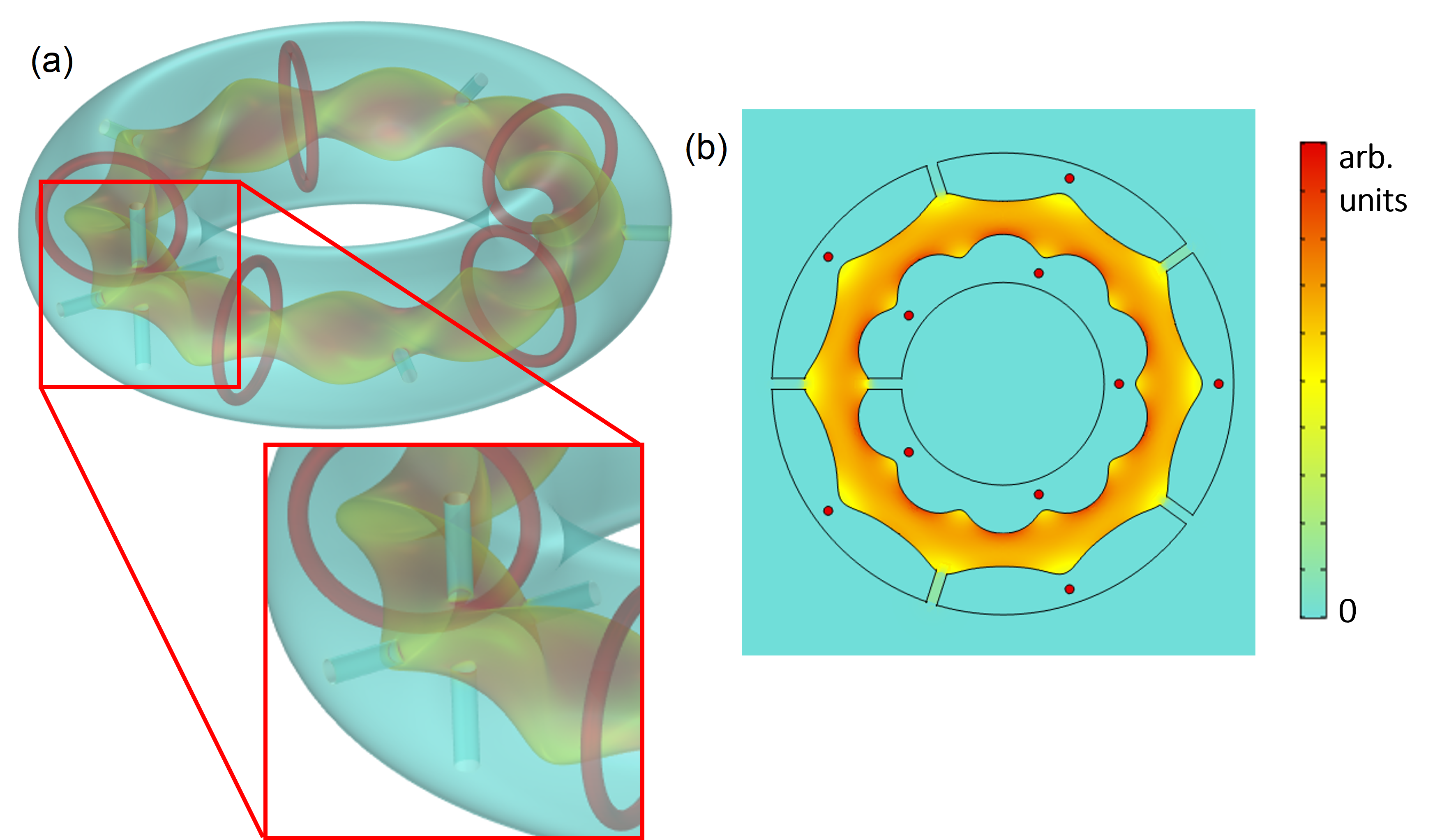}
\caption{Finite-element simulations of the 3D color map of the magnetic-field strength created by a superconducting toroid with a toroidal cavity. A total of eight holes had been drilled, one between each current loop except for one place were four holes had been drilled in the same transverse plan between to current loops. A zoom of the place were the four holes in the same transverse plane is provided. It can be observed that \textbf{B} does not exit outside the cavity through any hole. (b) is a horizontal cut plane of (a).}
\label{fig:LHD_plane}
\end{figure}

Another way of adding more holes to this kind of system is by placing a number of them in a transverse plane, as previously explained. This strategy allows us to access to the plasma through more holes and without increasing the number of coils, preserving the magnetic flux surfaces of the cavity and avoiding field leakage (see Fig. \ref{fig:LHD_plane}).

To sum up, in order to access the inner cavity there are several possible alternatives, either by playing with the size of the holes or by increasing the number of holes. In practice a smart combination of all strategies would be the best way to access the plasma while maintaining flux surfaces and minimizing field leakage.

\end{document}